\documentclass[aps,prb, twocolumn, ,superscriptaddress,showpacs,floatfix, letterpaper]{revtex4-1}
\usepackage[english]{babel}
\usepackage{graphicx}
\usepackage{stmaryrd}
\usepackage{amssymb}
\usepackage{amsfonts}
\usepackage{amsmath}
\usepackage{color}
\usepackage{xspace}

\usepackage{bm}

\begin{document}

\title{Electron spin resonance shifts in $S=1$
antiferromagnetic chains}
\author{Shunsuke C. Furuya}
\affiliation{DPMC-MaNEP, University of Geneva, 24 Quai Ernest-Ansermet CH-1211 Geneva, Switzerland}
\author{Yoshitaka Maeda}
\affiliation{Analysis Technology Center, Fujifilm Corporation,
Kanagawa 250-0193, Japan}
\author{Masaki Oshikawa}
\affiliation{Institute for Solid State Physics, University of Tokyo,
Kashiwa 277-8581, Japan}
\date{\today}
\begin{abstract}
We discuss electron spin resonance (ESR) shifts in spin-1
Heisenberg antiferromagnetic chains with a weak single-ion anisotropy,
based on several effective field theories:
the O(3) nonlinear sigma model (NLSM) in the Haldane phase,
free fermion theories around the lower and the upper critical fields.
In the O(3) NLSM, 
the single-ion anisotropy corresponds to a composite operator
which creates two magnons at the same time and position.
Therefore, even inside a parameter range where free magnon
 approximation is valid for thermodynamics, 
we have to take interactions among magnons into account in order to
 include the single-ion anisotropy as a perturbation.
Although the O(3) NLSM is only valid in the Haldane phase,
an appropriate translation of Faddeev-Zamolodchikov operators of the
 O(3) NLSM to fermion operators enables one to treat ESR shifts near the
 lower critical field in a similar manner to discussions in the Haldane
 phase.
 Our theory gives
quantitative agreements with a numerical evaluation using Quantum Monte
 Carlo simulation, and also with recent ESR experimental results on a
 spin-1 chain compounds NDMAP.
\end{abstract}
\maketitle

\section{\label{sec:introduction}Introduction}

Quantum phase transition has been studied for a long time.
In quantum magnetism, the magnetic field is the most familiar parameter
to cause quantum phase transitions.
An $S=1$ Heisenberg antiferromagnetic (HAF) chain and an $S=1/2$ two-leg
HAF ladder are typical examples of one-dimensional quantum spin systems
which show quantum phase transitions induced by the magnetic field.
These systems have the unique ground state separated from  excited
states by a finite excitation gap, at zero field.
As the magnetic field is gradually applied,
the excitation gap  is going to
vanish.~\cite{affleck_bec_Haldane,chitra_ladder}
After the collapse of the excitation gap,
the system enters into a field-induced critical phase.
The field-induced critical phase lies in a range $H_{c1} < H < H_{c2}$.
Here $H_{c1}$ and $H_{c2}$ are called as a lower and an upper critical
field.
For $H < H_{c1}$, the system is in the gapped phase.
And for $H_{c2} < H$, the system is
in another gapped phase where the spins are
fully polarized.
$H_{c2}$ is also called as a saturation field.

The quantum phase transitions at $H=H_{c1}$ and $H=H_{c2}$
bring about reconstructions of the excitation spectrum.
Especially, dynamical properties of low-energy excitations are
dramatically changed.
Recently, dynamics of electron spins in the field-induced critical phases
are actively investigated by various experimental
techniques.~\cite{Klanjsk_BPCB,cizmar_BPCB,kashiwagi_ndmap} 
Among these experimental techniques, 
electron spin resonance (ESR) occupies a unique position in its sensitivity
to interactions between electron spins.
In fact, thanks to this advantage of ESR, 
many interesting ESR experiments have been performed in one-dimensional
quantum spin
systems under high magnetic
field.~\cite{cizmar_BPCB,kashiwagi_ndmap,glazkov_NTENP}
These recent precise ESR experiments highlights necessity of
reliable quantitative theory of ESR in the field-induced critical phase.

Despite the theoretical and experimental importance of 
the field-induced critical phase, ESR in the field-induced critical
phase is less studied by theorists.
This situation is in contrast to the fact that $S=1/2$ HAF critical
chain whose low-temperature ESR is well
understood.~\cite{oshikawa_esr,brockmann_ESR_PRL,brockman_ESR_PRB,maeda_shift}
Although ESR of the $S=1$ HAF chain has been 
studied in several works, they were mostly concerned with ESR in gapped
phases.~\cite{affleck_ESR_Haldane,sakai_nenp}
It is the objective of the present paper to fill this gap by developing
 a theory of ESR in the field-induced critical phase, especially around
quantum critical points, of one-dimensional quantum spin systems
in an organized manner.

In this paper, we consider an $S=1$ HAF
chain with a general form of a single-ion anisotropy 
\begin{align}
 \mathcal H 
 &= J \sum_j \bm S_j \cdot \bm S_{j+1} - g_e\mu_B H\sum_j S^z_j \notag \\
 & \qquad + D\sum_j (S^p_j)^2 +E\sum_j \bigl[(S^q_j)^2 - (S^r_j)^2\bigr]
 \label{eq:H}
\end{align}
in the whole range of the magnetic field, from zero field $H=0$ to the 
saturation field $H=H_{c2}$. 
$p,q$ and $r$ refer to the principal axes of the single-ion anisotropy.
$g_e$ and $\mu_B$ are Land\'e $g$ factor of electron spin and
$\mu_B$ is Bohr magneton.
We put $\hbar = k_B = g_e \mu_B = 1$ unless otherwise stated.
In particular, we focus on a shift of the resonance frequency (ESR shift) 
caused by weakly anisotropic spin-spin interactions.

We reported, in our preceding Rapid Communication,~\cite{furuya_ffpt}
that the ESR shift in the range $H \lesssim H_{c1}$
is well explained by, the so-called form factor perturbation theory~\cite{controzzi_ffpt} (FFPT)
around an integrable field theory.
In the case of $S=1$ HAF chain, the O(3) nonlinear sigma model (NLSM)
plays the role of the unperturbed integrable field theory in FFPT.
In the Rapid Communication,~\cite{furuya_ffpt}
we applied the FFPT to the analysis of the ESR shift in $H \approx 0$ and
$ H_{c1}$, where
we utilized a close relation of effective field theories in two
different regions, $H \approx 0$ and $H \approx H_{c1}$.
This paper is also intended to take a closer look at this remarkable
feature.

In the next section, we will briefly review a general framework of
perturbative treatments for the ESR shift.
We consider ESR shifts in three regions: the low-field gapped region
(Sec.~\ref{sec:gapped}); the region near the lower critical field
(Sec.~\ref{sec:Hc1}); and the region near the upper critical field
(Sec.~\ref{sec:Hc2}).
In each region, we introduce an effective field theory and apply it to
the analysis of the ESR shift at low temperature.
Sec.~\ref{sec:ndmap} is devoted to a comparison of our theory with
recent ESR experiments~\cite{kashiwagi_ndmap} of the $S=1$ HAF compound NDMAP.
In Appendix~\ref{app:exchange}, we  discuss a qualitative difference
of the single-ion anisotropy and an exchange anisotropy from the
viewpoint of ESR shifts.

\section{\label{sec:shift}Framework}

Here we briefly review the perturbation theory of the ESR shift.
ESR experiments measure an absorption of an electromagnetic wave
by electron spins, where a microwave is typically applied.
From the absorption spectrum, we are able to extract information on
dynamics of electron spins.
Within the linear response theory, the ESR spectrum $I(\omega) \propto
\omega \chi''_{+-}(\omega, q)$ is written in terms of the retarded
Green's function,
\begin{equation}
 \chi''_{+-} (\omega, q=0) = \operatorname{Im}\biggl[i \int_{0}^\infty dt \,
 e^{i\omega t} \, \langle [S^+(t), S^-(0)] \rangle\biggr].
 \label{eq:chi}
\end{equation}
Here $S^\pm = S^x \pm i S^y$ denote transverse components
of the total spin $\bm S = \sum_j \bm S_j$, which is the generator of
the global SU(2) symmetry. 
Thus, if the whole Hamiltonian preserves the SU(2) symmetry in the spin
space, \eqref{eq:chi} is trivially constant.
In the presence of the magnetic field, the symmetry of the Hamiltonian
is lowered to U(1) at most.
If spin-spin interactions preserve the SU(2) symmetry,
Eq.~\eqref{eq:chi} is  still simple despite the presence of
interactions. 
\begin{equation}
 \chi''_{+-}(\omega, q=0) = 2\pi \langle S^z\rangle \delta(\omega -
  H).
  \label{eq:chi_SU(2)}
\end{equation}
The resonance frequency $\omega_r$
equals to the paramagnetic one $\omega_r = H$ at any
temperature.

If spin-spin interactions do not preserve the SU(2) symmetry,
the above discussion breaks down and the resonance frequency is shifted
from the paramagnetic one.
Let us assume that the Hamiltonian is composed of the three terms:
\begin{equation}
 \mathcal H = \mathcal H_{0} + \mathcal H_Z + \mathcal H',
  \label{eq:H_3}
\end{equation}
where $\mathcal H_0$ represents SU(2) symmetric interactions,
$\mathcal H_Z$ is the Zeeman term, and $\mathcal H'$ represents
anisotropic interactions.
The model \eqref{eq:H} falls into the form of Eq.~\eqref{eq:H_3}.
If the anisotropic interaction is weak, we are able to
consider a perturbative expansion of the resonance frequency  in the
anisotropy $\mathcal{H}'$. 

The first order perturbative expansion of the resonance frequency
was proposed first by Kanamori and Tachiki~\cite{KanamoriTachiki62} and later
applied to quantum spin systems
by Nagata and Tazuke.~\cite{nagata-tazuke_epr,nagata_EPR,maeda_shift}
Ref.~\onlinecite{maeda_perturbation} derived the ESR shift
$\delta \omega = \omega_{ r} - H$ 
from equal-time correlations at the lowest order in a general formalism,
\begin{equation}
 \delta \omega = -\frac{\langle [[\mathcal H', S^+], S^-]
  \rangle_0}{2\langle S^z \rangle_0} + \cdots.
  \label{eq:NT}
\end{equation}
The average $\langle \cdots \rangle_0$ is taken with respect to the
unperturbed Hamiltonian $\mathcal H^{(0)}$,
\begin{equation}
 \mathcal H^{(0)} = \mathcal H_0 + \mathcal H_Z.
  \label{eq:H0}
\end{equation}

While we thus far treated the ESR spectrum as a function of the
frequency $\omega$ with a fixed $H$ in above discussions,
this is often not the case in actual ESR experiments.
The ESR spectrum is usually obtained as a function of $H$
with a fixed $\omega$.
In this case, the ESR shift is defined as
\begin{equation}
 \delta H =  H_r - \omega/g_e\mu_B.
  \label{eq:dH}
\end{equation}
$H_r$ is the resonance field.
Note that the $g$ factor used in \eqref{eq:dH} is determined at
the high temperature limit.
By definition, \eqref{eq:dH} approaches zero as $T \to + \infty$.
At a low temperature $T \lesssim J$, it generaly holds that $\delta H
\not=0$.
According to Refs.~\onlinecite{brockmann_ESR_PRL,brockman_ESR_PRB},
within the first order accuracy, the ESR shift \eqref{eq:dH} satisfies
\begin{equation}
 g_e\mu_B \delta H = \frac{\langle[[\mathcal H', S^+], S^-] \rangle_0}{2\langle
  S^z \rangle_0}.
  \label{eq:NT_H}
\end{equation}
We should emphasize that Eq.~\eqref{eq:NT_H} 
is equivalent to \eqref{eq:NT}.
Therefore, as long as we are concerned with the first order perturbation
theory around \eqref{eq:H0}, 
it does not matter whether we change $\omega$ or $H$.

We apply the formula \eqref{eq:NT} to our model \eqref{eq:H}, namely,
\begin{align}
 \mathcal H^{(0)} &= J\sum_j \bm S_j \cdot \bm S_{j+1}
  -\sum_j \bm H \cdot \bm S_j, 
  \label{eq:H_0th} \\
 \mathcal H' &= D\sum_j (S^c_j)^2 + E\sum_j \bigl[(S^a_j)^2
 -(S^b_j)^2\bigr].
 \label{eq:SIA}
\end{align}
The ESR shift \eqref{eq:NT} is, in this case, given by
\begin{align}
 \delta \omega
 &= f_D(\bm z)Y_D(T,H),
 \label{eq:dw_D} \\
 f_D(\bm z)
 &= D(1-3{z_c}^2) -3E ({z_a}^2 - {z_b}^2),
 \label{eq:fD} \\
 Y_D(T,H)
 &=\frac 1{2\langle S^z \rangle_0} \sum_j \bigl[3\langle
 (S^z_j)^2\rangle_0 - 2 \bigr]
 \label{eq:YD}
\end{align}
The unit vector $\bm z \equiv \bm H/H$ is parallel to the magnetic field.
$z$ is represented as $\bm z = (z_a, z_b, z_c)$ in
the principal $(a,b,c)$ coordinate in \eqref{eq:H}.
$f_D(\bm z)$ is a constant (independent of $T$ and $H$) if the orientation
of the magnetic field is fixed.
For simplicity, we hereafter set $\bm H = H\hat z$ where $\hat z$ is the
unit vector along the $c$ axis, that is $\hat z = (0,0,1)$ in the principal axis
coordinate. 
We call $Y_D(T,H)$ as a normalized ESR shift.
The normalized ESR shift is useful for our purpose because
it can be applied to systems with any value of $D$ and $E$.

\section{\label{sec:qmc}QMC results}

\begin{figure}
 \centering
 \includegraphics[width=\linewidth]{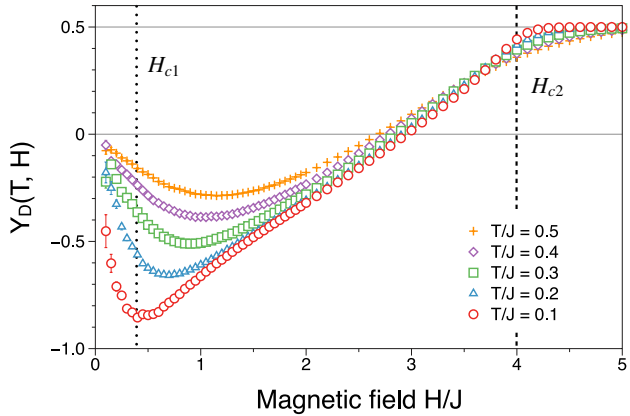}
 \caption{(Color online) Quantum Monte Carlo results of the normalized
 ESR shift~\eqref{eq:YD} induced by the single-ion anisotropy \eqref{eq:SIA}
 for temperatures $T/J = 0.1$ -- $ 0.5$.
 The system size is $L=40$ sites.
 The lower critical field $H_{c1} = 0.41J$ and the upper critical field
 $H_{c2} = 4J$ are guided by the dotted and the dashed lines
 respectively.
 There is an extremum around $H =H_{c1} + T$.
 This non-monotonic behavior of the ESR shift is understood by the
 finite temperature crossover.}
 \label{fig:qmc}
\end{figure}

We numerically evaluate the normalized ESR shift \eqref{eq:YD}
from quantum Monte Carlo (QMC) calculations.
The QMC results of the $H$ dependence of $Y_D(T,H)$ is shown in
Fig.~\ref{fig:qmc}.
We find several characteristics from Fig.~\ref{fig:qmc}.
(i) The normalized shift is approximately proportional to $H$
in the ranges $0<H<H_{c1}$ and $H_{c1} < H < H_{c2}$.
The slope $\partial_H Y_D(T,H)$ is negative in the former and positive
in the latter range.
(ii) The normalized shift has a minimum around $H=H_{c1}$.
The field which gives the minimum increases as the temperature
increases.
(iii) The normalized shift becomes zero at a certain value of $H$ because
$Y_D(T,H=0)=0$ and $\partial_H Y_D(T,H)|_{H=0} < 0$ hold at $H=0$
and the saturating value $Y_D(T,H>H_{c2})$ is positive.
Note that the field dependence in Fig.~\ref{fig:qmc} is
qualitatively different from that of $S=1/2$ HAF two-leg ladder
systems.~\cite{furuya_BPCB}
In $S=1/2$ HAF two-leg ladder systems, we fail to find
the change of the sign of the ESR shift.
Namely, the normalized shift is non-zero
from the infinitely weak field to the saturation field  (see Fig.1 in
Ref.~\onlinecite{furuya_BPCB}). 
The above three features suggest that the magnetic field dependence of
the normalized ESR shift reflects the finite-temperature
crossover.
In the following, we analyze the normalized shift \eqref{eq:YD} in the gapped,
lower critical, and upper critical regions.

\section{\label{sec:gapped}Low-field gapped phase}

\subsection{Effective field theory}

First we review the zero-field case, then we will extend the argument to
the low-field case.
The unperturbed model \eqref{eq:H0} in the absence of the magnetic
field,
\begin{equation}
 \mathcal H_0 =J \sum_j \bm S_j \cdot \bm S_{j+1}.
  \label{eq:H_zeroH}
\end{equation}
has an excitation gap $\Delta_0 = 0.41J$,~\cite{Todo} which is called as the
Haldane gap.
Haldane proposed that HAF chains with an integer quantum spin number $S$
have an excitation gap $\Delta_0$ based on a semiclassical field theory,
the O(3) nonlinear sigma model
(NLSM).~\cite{haldane-conjecture_1,haldane-conjecture_2}
The O(3) NLSM has a Lagrangian,
\begin{equation}
 \mathcal L = \frac 1{2g}\partial_\mu \bm n \cdot \partial^\mu \bm n +
  \frac{\Theta}{4\pi} \bm n \cdot \partial_t \bm n \times \partial_x \bm n.
  \label{eq:A_NLSM}
\end{equation}
The contraction $\partial_\mu \bm n \cdot \partial^\mu \bm n =
(\partial_t \bm n)^2 - (\partial_x \bm n)^2$ was taken.
For simplicity we put the spin-wave velocity to unity.
The field $\bm n(t, x)$ represents an antiferromagnetic order:
\begin{equation}
 \bm S_x \sim \sqrt{S(S+1)} (-1)^x \bm n(x)+ \bm L(x)
  \label{eq:S2nL}
\end{equation}
The uniform component $\bm L =  \bm n \times \partial_t \bm n/g$ 
is quadratic in $\bm n$.
The coupling constant $\Theta = 2\pi S$ is equal to $0$ or $\pi$ mod
$2\pi$.

The O(3) NLSM is integrable when $\Theta \equiv 0, \pi$ (mod $2\pi$).
In the case $\Theta \equiv \pi$, the O(3) NLSM is
critical.~\cite{AffleckHaldane}
On the other hand, in the case $\Theta \equiv 0$ of our interest,
the O(3) NLSM has massive triplet particles, which is called magnons,
as the lowest excitations.
The triplet magnons are created
by $ n^a(t,x)\propto (-1)^x S^a(t,x)$ ($a=x,y,z$).
Thus, the field $S^a$ satisfies the relation
\begin{equation}
 (-1)^x\langle 0|S^a(t,x)|\theta_1,a_1 \rangle =  \delta_{aa_1}\sqrt{Z}
  e^{ix^\mu p_\mu}.
  \label{eq:ff_n}
\end{equation}
$|0\rangle$ is the ground state, $|\theta_1, a_1\rangle$ is a
one-magnon state with the rapidity $\theta_1$ and the index
$a_1 = x,y,z$, and $Z$ is the renormalization factor which will
be discussed later.
The O(3) NLSM is Lorentz invariant, and the triplet excitations obey
a dispersion relation $p_0 = \sqrt{{\Delta_0}^2 +
{p_1}^2}$ parametrized by a single parameter $\theta$.
\begin{equation}
 p_0 = \Delta_0 \cosh \theta, \quad p_1 = \Delta_0 \sinh \theta
  \label{eq:p0p1}
\end{equation}
This parameter $\theta$ is called as a rapidity, which uniquely determines
the energy $p_0$ and the momentum $p_1$ of magnons.
Therefore, the one-magnon state $|\theta_1, a_1 \rangle$ is fully
characterized by the rapidity $\theta_1$ and the index $a_1$.
We normalize the state $|\theta, a\rangle$ by
\begin{equation}
 \langle \theta_1, a_1|\theta_2, a_2\rangle = 4\pi
  \delta_{a_1a_2}\delta(\theta_1 -   \theta_2).
\end{equation}
$n$-magnon states, $|\theta_1,a_1; \cdots; \theta_n,a_n\rangle$ are
specified by a set of rapidities $\{\theta_1, \cdots, \theta_n\}$
and indices $\{a_1, \cdots, a_n\}$.
They are normalized as follows:
\begin{multline}
 \langle \theta_1, a_1; \cdots; \theta_n, a_n|\theta'_1, a'_1;
 \cdots; \theta'_n, a'_m \rangle \\
  = \delta_{nm}(4\pi)^n \prod_{l=1}^n
 \delta_{a_la'_l}\delta(\theta_l- \theta'_l)
\end{multline}
A matrix element
\begin{equation}
 F_{\mathcal O}(\theta_1, a_1; \cdots; \theta_n, a_n)
  = \langle 0|\mathcal O(0)|\theta_1, a_1; \cdots; \theta_n, a_n\rangle
  \label{eq:ff_def}
\end{equation}
is called as an $n$-magnon form factor of a local operator $\mathcal
O(t,x)$.
Here  $\mathcal O(0)$ is an abbreviation of $\mathcal O(0,0)$.
A Lorentz boost of the O(3) NLSM alters \eqref{eq:ff_def} to
\begin{align}
 &\langle 0|\mathcal O(t,x)|\theta_1, a_1; \cdots; \theta_n, a_n\rangle
 \notag \\
 & \qquad =F_{\mathcal O}(\theta_1, a_1; \cdots; \theta_n, a_n)
 e^{i(tP_0-xP_1)},
\end{align}
where $P_0$ and $P_1$ denote the total energy and momentum.
\begin{equation}
 P_0 = \sum_{m=1}^n \Delta_0 \cosh \theta_m, \quad
  P_1 = \sum_{m=1}^n \Delta_0 \sinh \theta_m.
\end{equation}

For example, the relation~\eqref{eq:ff_n} is equivalent to the
one-magnon form factor of $S^a$ at the origin
\begin{equation}
 F_{S^a}(\theta_1,a_1) = (-1)^r \sqrt{Z} \delta_{a,a_1}.
\label{eq.ff1p}
\end{equation}
The relation \eqref{eq:ff_n} connects the low-energy effective field
theory and the physical operator $S^a(t,x)$ in the original spin
model.
The renormalization factor $\sqrt{Z}$
inevitably depends on short-distance, non-universal
physics and cannot be determined within the effective field theory.
$Z$ is determined only by numerical calculations.
$Z\approx 1.26$ is concluded from
ensity matrix renormalization group
calculations.~\cite{sorensen_Sk,sorensen_haldane} 
It is emphasized that Eq.~\eqref{eq.ff1p} should \emph{not}
be interpreted as an identity between
the physical spin operator $S^a$ and a creation operator of magnons.
The spin operator $S^a$ also has nonvanishing higher-order form factors.
Thus the form factor of the powers of $S^a$ is not solely
determined by the one magnon form factor~\eqref{eq.ff1p},
even in the leading order.

Let us consider the traceless symmetric tensor
\begin{equation}
 \Sigma^{ab} \equiv S^aS^b- \frac 23\delta_{ab} .
  \label{eq:Sigma}
\end{equation}
$\Sigma^{ab}$ has a two-magnon form factor,
\begin{equation}
 F_{\Sigma^{ab}}(\theta_1, a_1; \theta_2, a_2)
  = -iZ_2 \delta_{ab}\delta_{a_1a_2}(3\delta_{aa_1}-1)\psi_2(\theta_1 -
  \theta_2).
  \label{eq:ff_Sigma}
\end{equation}
In the case of O($N$) NLSM,~\cite{balog_formfactor} $\psi_2(\theta)$ 
is given by an integral.
\begin{align}
 \psi_2(\theta) 
 &= \sinh\biggl( \frac \theta 2\biggr)
  \exp \biggl[\int_0^\infty \frac{dx}x K_N(x)
  \frac{\cosh[(\pi + i \theta)x]-1}{\sinh (\pi x)}\biggr]
  \label{eq:psi2_O(N)} \\
 K_N(x) 
 &= \frac{e^{-\pi x} + e^{-2\pi x/(N-2)}}{1+e^{-\pi x}}.
\end{align}
We performed the integral and derived an explicit form of
$\psi_2(\theta)$ for the $N=3$ case in our preceding paper.~\cite{furuya_ffpt}
\begin{equation}
 \psi_2(\theta) = \frac i2(\theta - \pi i) \tanh \biggl( \frac \theta
  2\biggr)
  \label{eq:psi2_O(3)}
\end{equation}
The two-magnon form factor \eqref{eq:ff_Sigma} is now determined
except for the non-universal factor $Z_2$.
We emphasize that $Z_2$ is an independent parameter from $Z$.
We have determined $Z_2 \approx 0.24$ by comparing the NLSM prediction
with the correlation function of $(S^a)^2$ obtained numerically using
the infinite time evolving block decimation method.~\cite{furuya_ffpt}

The basis $\{|\theta_1, a_1; \cdots; \theta_n, a_n\rangle \}$ with $n=0,1,2,
\cdots$ is complete and orthonormal.
The identity $\hat 1$ reads as
\begin{align}
 \hat 1 &=|0\rangle \langle 0| + \sum_{n=1}^\infty \frac 1{n!}
 \sum_{a_1 \cdots a_n}\int_{-\infty}^\infty \frac{d\theta_1 \cdots
 d\theta_n}{(4\pi)^n} \notag \\
 & \qquad \times |\theta_1,a_1 ; \cdots; \theta_n, a_n \rangle
 \langle \theta_1, a_1; \cdots; \theta_n, a_n|.
 \label{eq:1}
\end{align}

We note an important relation of form factors,
the crossing relation.
In subsequent sections, we will encounter matrix elements such as
$\langle \theta_2, a_2|\mathcal O(0)|\theta_1, a_1 \rangle$.
The crossing relation allows one to relate this matrix element
to form factors.
\begin{align}
 \langle \theta_2, a_2|\mathcal O(0)|\theta_1, a_1\rangle
 &= \langle 0|\mathcal O(0)|\theta_1,a_1; \theta_2 - \pi i,
 \bar{a_2}\rangle \notag \\
  &= F_{\mathcal O}(\theta_1, a_1; \theta_2 - \pi i, \bar{a_2})
  \label{eq:crossing}
\end{align}
The index $\bar{a}$ represents an index of an anti-magnon
conjugate to the magnon with the index $a$.
If we employ the labeling $a=x,y,z$, then $\bar a = a$ holds.
If, on the other hand, we employ a labeling $a=+,0, -$, namely $(n^+, n^0, n^-) 
= ((n^x+in^y)/\sqrt{2}, n^z, (n^x-in^y)/\sqrt 2)$, we have $\bar + = -$, $\bar 0 = 0$ and
$\bar - = +$.

Under a weak magnetic field $H < \Delta_0$,
the unperturbed system \eqref{eq:H_0th} still has a finite gap $\Delta_0 - H$.
Here we have to replace the dispersion relation \eqref{eq:p0p1} to
\begin{equation}
 p_0 = \Delta_0 \cosh \theta - aH, \quad
  p_1 = \Delta_0 \sinh \theta,
\end{equation}
where $a=0, +, -$.
Namely, the triplet degeneracy is lifted by the Zeeman splitting term.
If the magnetic field is very weak $H \ll \Delta_0$, then
we may use the form factors evaluated for the $H=0$ case at the lowest
order of $H$.
 For this purpose, in the following, we use the labeling $a=+, 0, -$ of
 magnons, which corresponds to energy eigenstates under the magnetic
 field.

\subsection{ESR shift}

In the limit $H, T \to 0$, the density of magnons becomes low.
It should be reasonable in this dilute limit that we ignore
contributions of multi-magnon states to thermodynamic quantities, for
instance, the magnetization and the normalized shift \eqref{eq:YD}.
We multiply a projection operator
\[
 P_1 = \sum_{a=0,+,-}\int_{-\infty}^\infty \frac{d\theta}{4\pi}\,
 |\theta, a \rangle\langle \theta, a|
\]
to an operator $\mathcal O$
so that the multi-magnon contributions to the average
$\langle \mathcal O \rangle$ are projected out.
Let us consider $\mathcal O = \Sigma^{00}(0,x)$.
Using the crossing relation \eqref{eq:crossing}
and the two-magnon form factor \eqref{eq:ff_Sigma}, we obtain
\begin{align}
 &P_1 \Sigma^{00}(0,x) P_1 \notag \\
 &\quad =-iZ_2 \int_{-\infty}^\infty \frac{d\theta d\theta'}{(4\pi)^2}
 \psi_2(\theta' - \theta + \pi i) e^{ix[P_1(\theta') -P_1(\theta)]}
 \notag \\
 & \qquad \times \bigl(2|\theta, 0\rangle\langle \theta, 0|
 -|\theta, +\rangle \langle \theta', +|
 -|\theta, -\rangle \langle \theta', -|\bigr).
 \label{eq:P1SigmaP1}
\end{align}
Thus, in the dilute limit,
the numerator $\sum_j [3\langle (S^z_j)^2\rangle_0-2]$ of
the normalized shift \eqref{eq:YD} is approximated as follows.
\begin{align}
 \sum_j& \bigl[ 3\langle (S^z_j)^2\rangle_0 - 2\bigr] \notag \\
 &= 3\int dx\,  \langle \Sigma^{00}(0,x) \rangle_0 \notag \\
 &\sim- 6Z_2 \int_{-\infty}^\infty \frac{vd\theta}{4\pi
 E(\theta)}\,  e^{-E(\theta)/T}\sinh^2 \biggl( \frac H{2T}\biggr)
 \label{eq:num_Y_dilute}
\end{align}
Here $E(\theta) = \Delta_0 \cosh \theta$ is the zero-field dispersion.
Similarly, the magnetization is given by
\begin{equation}
 \langle S^z \rangle_0
  \sim 2\sinh\biggl( \frac HT\biggr) \int_{-\infty}^\infty
  \frac{d\theta}{4\pi} 
  e^{-E(\theta)/T}.
  \label{eq:mag_dilute}
\end{equation}
From \eqref{eq:num_Y_dilute} and \eqref{eq:mag_dilute}, the 
normalized shift in the dilute limit reads
\begin{equation}
 Y_D(T,H) = -\frac{3Z_2}4\tanh \biggl(\frac H{2T}\biggr)
  \frac{\displaystyle\int_{-\infty}^\infty \frac{vd\theta}{4\pi
  E(\theta)} e^{-E(\theta)/T}}{\displaystyle\int_{-\infty}^\infty
  \frac{d\theta}{4\pi } e^{-E(\theta)/T}}.
  \label{eq:YD_dilute}
\end{equation}
Eq.~\eqref{eq:YD_dilute} correctly reproduces the features of the normalized
shift, $Y_D(T,H) \propto H$ and $\partial_H Y_D(T,H) < 0$ in the
limit $H \to 0$. 
However, \eqref{eq:YD_dilute} cannot explain the upturn of the
normalized shift around $H=H_{c1}$.
In order to extend \eqref{eq:YD_dilute} to the region $H \sim H_{c1}$,
we must take into account multi-magnon states.

\section{\label{sec:Hc1}Near lower critical field}

\subsection{\label{sec:fermion}Effective field theory}

At $H=\Delta_0$, 
the lowest magnon band specified by the index $a=+$ touches
the ground state.
The point $H_{c1} \equiv \Delta_0$ corresponds to a quantum critical
point.
Above $H_{c1}$, gapless excitations exist.
Thus, $H = H_{c1}$ separates the low-field gapped phase (called as the
Haldane phase) and the high-field gapless phase (the field induced
critical phase).
We call $H_{c1}$ the lower critical field.
The quantum phase transition occurs only at $T=0$. 
Nevertheless, at finite temperatures, in a range of magnetic field $H -
H_{c1} \leq T$, which is called as ``quantum critical region'', properties
of the system reflect the nature of the quantum critical
point.~\cite{giamarchi_nature}

It is  known that a free fermion theory 
describes low-energy behavior of $S=1$ HAF chain in the quantum
critical
region.~\cite{Lieb63a,Lieb63b,affleck_GL_haldane,affleck_bec_Haldane,maeda_magnetization}
The free fermion has a dispersion relation,
\begin{equation}
 E(k) = \frac{k^2}{2\Delta_0} -\mu.
  \label{eq:Ek_critical}
\end{equation}
The chemical potential is $\mu = H- H_{c1}$.
As the chemical potential of the free fermion increases,
the number of the free fermion also increases.
In terms of spin systems, the number of the free fermion is
identical to the magnetization density $m_+(T,H) \equiv \langle S^z \rangle_0/L$:
\begin{align}
 m_+(T,H)
 &= \sqrt{\frac{\Delta_0}{2\pi^2}}\int_0^\infty d\epsilon\, D(\epsilon)
 f(\epsilon - \mu) \notag \\
 &=- \sqrt{\frac{T\Delta_0}{2\pi}}\operatorname{Li}_{1/2}(-e^{\mu/T})
 \label{eq:m+}
\end{align}
$L$ is the length of the spin chain,
$D(\epsilon) = \epsilon^{-1/2}$ is the density of states,
and $f(\xi) = (e^{\xi/T} + 1)^{-1}$ is the Fermi distribution function.
In the second line, the integral is performed explicitly,
with the result given in terms of the polylogarithm function
\begin{equation}
 \operatorname{Li}_n(x) = \sum_{m=1}^\infty \frac{x^m}{m^n}.
\end{equation}

Above the quantum critical region $H \gtrsim H_{c1}$,
a gapless excitation with a linear dispersion $E(k) \sim k$
dominates the low-temperature physics of the $S=1$ HAF chain.
The excitation is identified with the Tomonaga-Luttinger (TL)
liquid.~\cite{chitra_ladder,konik_TLL}
We do not go into detail on the TL liquid in the field-induced critical
phase. 

\subsection{ESR shift}

\begin{figure}
 \centering
 \includegraphics[width=\linewidth]{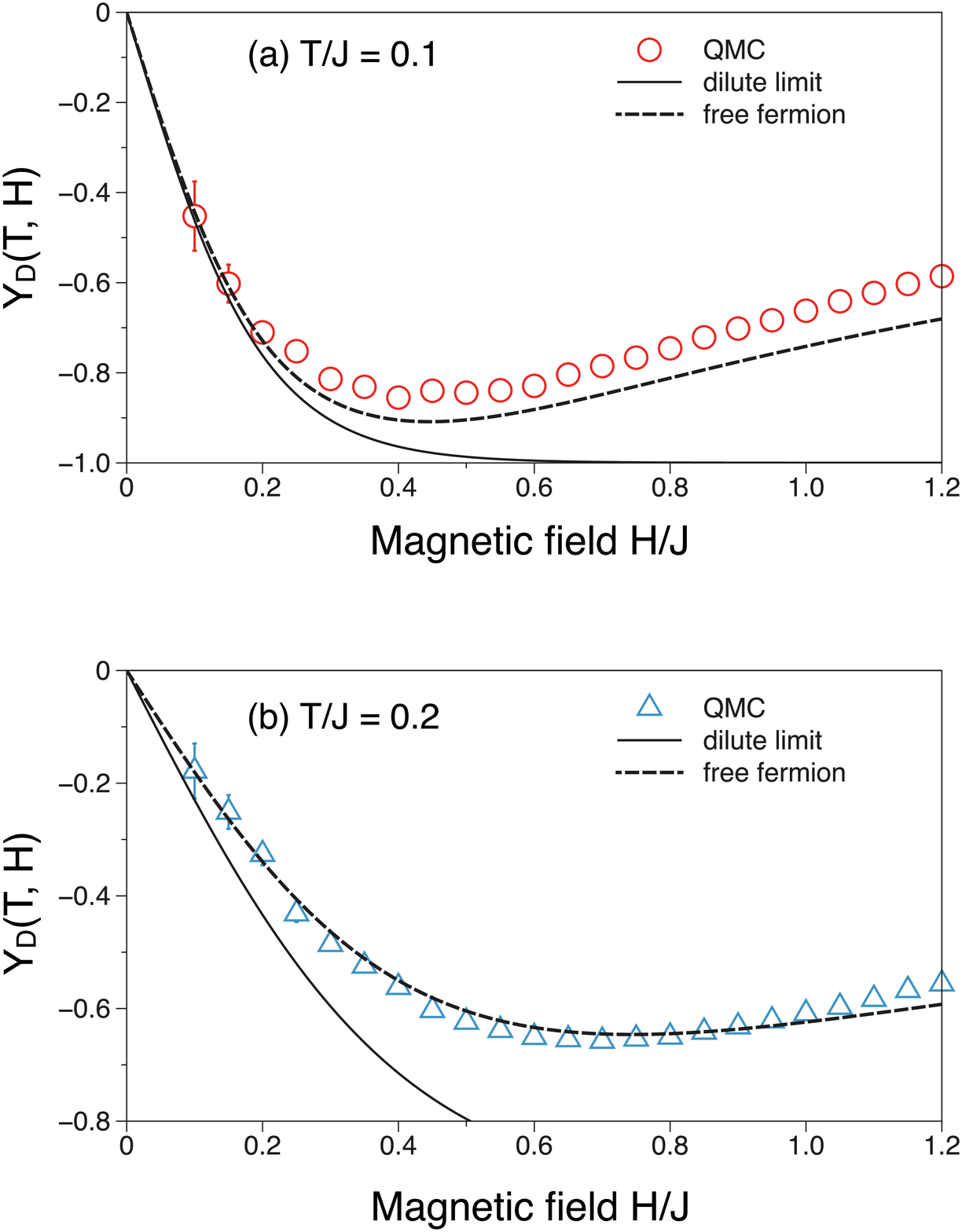}
 \caption{(Color online) Comparisons of QMC and analytic results
 at (a) $T/J = 0.1$ and at (b) $T/J=0.2$.
 Open symbols (circles and triangles) represents QMC data.
 The solid curves denote the normalized shift in the dilute limit
 \eqref{eq:YD_dilute}. 
 The dashed curves correspond to \eqref{eq:YD_Hc1}.
 }
 \label{fig:YD_Hc1}
\end{figure}

In the previous section, we formulated the O(3) NLSM with the
multi-magnon states, $|\theta_1, a_1; \cdots; \theta_n, a_n\rangle$.
Instead of these multi-magnon states, we may consider creation and
annihilation operators of magnons, which we denote $Z_a(\theta)$ and
$Z^\dagger_a(\theta)$ respectively.
Using them, we can create a one-magnon state
$|\theta, a \rangle$ and its conjugate,
\begin{equation}
 |\theta, a \rangle = Z^\dagger_a(\theta)|0\rangle,
  \quad
  \langle \theta, a|=\langle 0|Z_a(\theta).
\end{equation}
Similarly, the $n$-magnon state and its conjugate are given by
\begin{align*}
 |\theta_1, a_1; \cdots; \theta_n, a_n \rangle
 &= Z^\dagger_{a_1}(\theta_1) \cdots
 Z^\dagger_{a_n}(\theta_n)|0\rangle, \\
 \langle \theta_1, a_1; \cdots; \theta_n, a_n|
 &= \langle 0|Z_{a_n}(\theta_n)\cdots Z_{a_1}(\theta_1).
\end{align*}
These $Z_a(\theta)$ and $Z_a^\dagger(\theta)$ are called as
Faddeev-Zamolodchikov (FZ) operators and satisfy the following algebra.
\begin{align}
 Z_{a_1}(\theta_1) Z_{a_2}(\theta_2)
 &= S_{a_1a_2}^{b_1b_2}(\theta_1 - \theta_2)
 Z_{b_2}(\theta_2)Z_{b_1}(\theta_1)
 \label{eq:ZZ} \\
 Z^\dagger_{a_1}(\theta_1) Z^\dagger_{a_2}(\theta_2)
 &= S_{a_1a_2}^{b_1b_2}(\theta_1 - \theta_2)
 Z^\dagger_{b_2}(\theta_2)Z^\dagger_{b_1}(\theta_1)
 \label{eq:ZdZd} \\
 Z_{a_1}(\theta_1) Z^\dagger_{a_2}(\theta_2)
 &= 4\pi \delta_{a_1a_2}\delta(\theta_1 - \theta_2) \notag \\
 & \qquad + S_{a_2b_1}^{b_2a_1}(\theta_1 - \theta_2)
 Z^\dagger_{b_2}(\theta_2)Z_{b_1}(\theta_1)
 \label{eq:ZZd}
\end{align}
The factor $S_{ab}^{cd}(\theta)$ is an $S$ matrix.
The $S$ matrix possesses information of two-magnon scatterings.
If the magnon created by $Z^\dagger_a(\theta)$ were a free boson (a free
fermion),
the $S$ matrix would simply be $S_{ab}^{cd}(\theta) = \delta_{ad}\delta_{bc}$
($S_{ab}^{cd}(\theta) = -\delta_{ad}\delta_{bc}$).
In reality, the magnon is neither free boson nor fermion.
Thus, the $S$ matrix is a nontrivial function of the rapidity.
Fortunately $S$ matrix of the O(3) NLSM is exactly known.
\begin{equation}
 S_{ab}^{cd}(\theta) = \delta_{ab}\delta_{cd}\sigma_1(\theta)
  + \delta_{ac}\delta_{bd}\sigma_2(\theta) +
  \delta_{ad}\delta_{bc}\sigma_3(\theta) 
  \label{eq:Smatrix}
\end{equation}
$\sigma_i(\theta)$'s ($i=1,2,3$) are
\begin{align}
 \sigma_1(\theta) &= \frac{2\pi i \theta}{(\theta + \pi i)(\theta - 2\pi
 i)},
 \label{eq:sigma1} \\
 \sigma_2(\theta) &= \frac{\theta(\theta -\pi i)}{(\theta + \pi
 i)(\theta - 2\pi i)},
 \label{eq:sigma2} \\
 \sigma_3(\theta) &= \frac{2\pi i (\pi i - \theta)}{(\theta + \pi
 i)(\theta - 2\pi i)}.
 \label{eq:sigma3}
\end{align}

As well as the set of multi-magnon states $\{|\theta_1, a_1; \cdots;
\theta_n, a_n \rangle \}$, a set of FZ operators, $\{Z_a(\theta),
Z^\dagger_b(\theta')\}$ is complete.
In other words, we can expand the arbitrary operator $\mathcal O(t,x)$
in the power of FZ operators.
For instance, $\int dx \, \Sigma^{aa}(0,x)$ is expanded as
\begin{align}
 &\int dx \, \Sigma^{aa}(0,x)\notag \\
 &=\frac{Z_2}2 \int_{-\infty}^\infty \frac{vd\theta}{4\pi E(\theta)}
 \bigl[
 2Z^\dagger_0(\theta) Z_0(\theta) - Z_+^\dagger(\theta)Z_+(\theta)
 \notag \\
 & \quad -Z_-^\dagger(\theta)Z_-(\theta)\bigr]
 + (\mbox{higher-order terms}).
 \label{eq:Sigma2Z}
\end{align}
The omitted higher-order terms contain, for instance, a quartic term
$Z^\dagger_{a_1}(\theta_1)
Z_{a_2}(\theta_2)Z^\dagger_{a_3}(\theta_3)Z_{a_4}(\theta_4)$.
The projection \eqref{eq:P1SigmaP1} corresponds to an approximation
which drops the higher-order terms of \eqref{eq:Sigma2Z} out.
To improve the result \eqref{eq:YD_dilute},
we need to accurately evaluate the higher-order terms of the expansion
\eqref{eq:Sigma2Z}.

At low temperatures and around the lower critical field, we can focus on low-energy limit of NLSM. Here, the S-matrix of the O(3) NLSM actually simplifies as
\begin{equation}
S^{cd}_{ab}(\theta) \rightarrow -\delta_{ad}\delta_{bc},
\end{equation}
which is nothing but the S-matrix of free fermions. This implies that,
in this limit, we can replace the FZ operators by the fermion creation
and annihilation operators as
\begin{equation}
 Z_a(\theta) \sim \sqrt{\frac {2E(\theta)}{v}} \, c_a(k),
  \quad
 Z^\dagger_a(\theta) \sim \sqrt{\frac{2E(\theta)}{v}} \, c^\dagger_a(k),
 \label{eq:Z2c}
\end{equation}
with $k = \Delta_0 \sinh \theta$.
The rule \eqref{eq:Z2c} correctly reproduces the anticommutation
relations, 
\begin{align}
 \{c_a(k), c_{a'}(k') \}&=0,
 \label{eq:cc} \\
 \{c^\dagger_a(k), c^\dagger_{a'}(k')\} &= 0,
 \label{eq:cdcd} \\
 \{c_a(k), c^\dagger_{a'}(k')\} &= 2\pi \delta_{aa'} \delta(k-k'),
 \label{eq:ccd}
\end{align}
from Eqs.~\eqref{eq:ZZ}, \eqref{eq:ZdZd} and \eqref{eq:ZZd}.
This fermion has a dispersion $E_a(k) =
\sqrt{{{\Delta_0}^2} + k^2} - aH$ ($a=0,+,-$), and indeed corresponds
exactly to the free fermion effective theory discussed in Sec.~\ref{sec:fermion}. In
other words, the free fermion effective theory for the quantum critical
region is now derived systematically as a low-energy limit of the O(3)
NLSM under an applied field.


The replacement \eqref{eq:Z2c} enables us to compute
the normalized shift explicitly.
\begin{equation}
 Y_D(T,H) 
 = \frac{3Z_2}{2m(T,H)}\int_{-\infty}^\infty \frac{vdk}{4\pi E_0(k)}\,
  \bigl[2f_0(k) - f_+(k) - f_-(k)\bigr]
  \label{eq:YD_Hc1}
\end{equation}
$f_a(k) = (e^{(\sqrt{{\Delta_0}^2 + k^2} - aH)/T}+1)^{-1}$ is the Fermi
distribution function and $m(T,H)$ is the magnetization,
\begin{equation}
 m(T,H) = \int_{-\infty}^\infty \frac{dk}{2\pi}\, \bigl[f_+(k) -
  f_-(k)\bigr].
  \label{eq:mag_Hc1}
\end{equation}
The analytic result \eqref{eq:YD_Hc1} is compared with the QMC results
at $T = 0.1J$ and $0.2J$ in Fig.~\ref{fig:YD_Hc1}.
The free fermion representation \eqref{eq:YD_Hc1}
reproduces the minimum of the normalized ESR shift and, furthermore,
agrees quantitatively with the QMC data.
We stress that, the systematic derivation based on the exact form
factors of the O(3) NLSM is necessary to obtain Eq.~\eqref{eq:YD_Hc1} correctly. In
fact, it contains the nontrivial renormalization factor $Z_2$, which is
independent of the standard renormalization factor $Z$. A naive
application of the free fermion effective theory would lead to a formula
similar to Eq.~\eqref{eq:YD_Hc1} but with $Z$ appearing in the place of $Z_2$. Clearly, it
does not agree with the QMC result, demonstrating the importance of the
form-factor approach.

\section{\label{sec:Hc2}Near upper critical field}

\subsection{Effective field theory}

The field-induced critical phase ends at the upper critical field
 $H = H_{c2}$ where $H_{c2} =4J$.
Above the upper critical field, the spins are fully polarized, where
the gap opens again and the low-energy excitation has a parabolic
 dispersion.
Slightly below the upper critical field ($H_{c2} \ll H- H_{c2}<0$),
almost all spins are polarized.
Here we may neglect the $S^z_j = -1$ component anti-parallel to the
 magnetic field because it costs huge amounts of energy.
Thus, the $S=1$ spin is effectively described by an $S=1/2$ spin.
\begin{equation}
 S^z_j \sim \frac 12 (1+ \sigma_j^z), \qquad
 S^\pm_j \sim \frac 1{\sqrt 2}(-1)^j \sigma^\pm_j
 \label{eq:S1toS1/2}
\end{equation}
$(\sigma^x_j, \sigma^y_j, \sigma^z_j)$ is the Pauli matrices and
$\sigma^\pm_j \equiv (\sigma^x_j \pm i \sigma^y_j)/2$.
The unperturbed Hamiltonian \eqref{eq:H_0th} is transformed into
an effective $S=1/2$ XXZ chain.
\begin{equation}
 \mathcal H^{(0)}\sim \frac J2 \sum_j\biggl[- (\sigma^x_j \sigma^x_{j+1} +
  \sigma^y_j \sigma^y_{j+1}) +\frac 12 \sigma^z_j \sigma^z_{j+1}\biggr]
  -\frac h2 \sum_j \sigma^z_j
  \label{eq:H_0th_Hc2}
\end{equation}
This is effectively written in terms of a free fermion,
\begin{equation}
 \mathcal H^{(0)} \sim \int_{-\infty}^\infty \frac{dk}{2\pi} E(k)
  c^\dagger(k)c(k),
  \label{eq:freefermion_hc2}
\end{equation}
with a quadratic dispersion,
\begin{equation}
 E(k) = \frac{k^2}{2\mathfrak m} - \tilde \mu.
\end{equation}
Here $\mathfrak m = 1/2J$ and $\tilde \mu = H_{c2}-H$ are the mass of
the fermion and the chemical potential that the fermion feels.
Thus, the effective theories around the upper critical field field
and the lower critical field are isomorphic, while
the mass and the chemical potential of the fermions are different.
It should be also noted that the free fermion in each theory represents
a different object with respect to the original spin system.

\subsection{ESR shift}

\begin{figure}
 \centering
 \includegraphics[width=\linewidth]{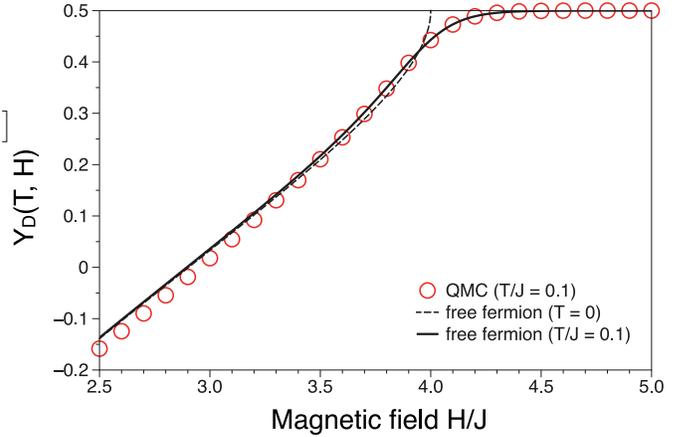}
 \caption{(Color online) The open circles denote the QMC data
 obtained for $40$-site chains at $T/J = 0.1$.
 The solid and dashed curves are derived from the free fermion theory
 near $H_{c2}$.
 The former is $T>0$ data and the latter is $T=0$ data.
 The finite-temperature effect is irrelevant in the lower field range
 $H \lesssim 3J$.
 }
 \label{fig:YD_Hc2}
\end{figure}

Using the mapping \eqref{eq:S1toS1/2}, one can represent the normalized
shift in the Pauli matrices.
\begin{equation}
 Y_D(T,H) = \frac 12 - \frac{1-\langle \sigma^z_j \rangle_0}{1+ \langle
  \sigma^z_j \rangle_0}
  \label{eq:YD_sigma}
\end{equation}
Here the average $\langle \cdots \rangle_0$ is taken by
the Hamiltonian \eqref{eq:H_0th_Hc2} of the effective
$S=1/2$ XXZ chain.
A free fermion theory with the dynamical exponent $z=2$ describes the
low-energy physics near the upper critical field $H_{c2}$.
Similarly to Eq.~\eqref{eq:m+}, 
the magnetization density $\langle \sigma^z_j \rangle_0$ is given by the
polylogarithm function as
\begin{equation}
 \langle \sigma^z_j \rangle_0 = 1+ 2\sqrt{\frac{T}{4\pi J}}
  \operatorname{Li}_{1/2}\bigl(-e^{(H_{c2}-H)/T} \bigr).
  \label{eq:mag_Hc2_T}
\end{equation}
Substituting \eqref{eq:mag_Hc2_T} into \eqref{eq:YD_sigma},
we obtain the explicit representation of the normalized shift.
We show the normalized ESR shift computed by the free fermion theory
in Fig.~\ref{fig:YD_Hc2}.
In order to see the field dependence explicitly, we consider the $T=0$ case.
The magnetization shows a singular dependence on the magnetic field at
$T=0$.
\begin{equation}
 \langle \sigma^z_j \rangle_0 = 1-\frac 2{\pi} \sqrt{H_{c2}-H} +
  \mathcal O(H_{c2}-H)
\end{equation}
The normalized shift at $T=0$ is shown by the dashed curve in
Fig.~\ref{fig:YD_Hc2}. 
The free fermion theory \eqref{eq:freefermion_hc2}
appears to work well in the entire region of Fig.~\ref{fig:YD_Hc2}
in the limit of $T \rightarrow 0$.
However, the numerical result in Fig.~\ref{fig:qmc} shows
a non-negligible temperature dependence for $H \lesssim 3J$
while the free fermion theory shows little temperature dependence.
This corresponds the breakdown of the present picture
based on spin flips from the saturated state, in the
lower magnetic field.

\section{\label{sec:ndmap}NDMAP}

\begin{figure}
 \centering
 \includegraphics[width=\linewidth]{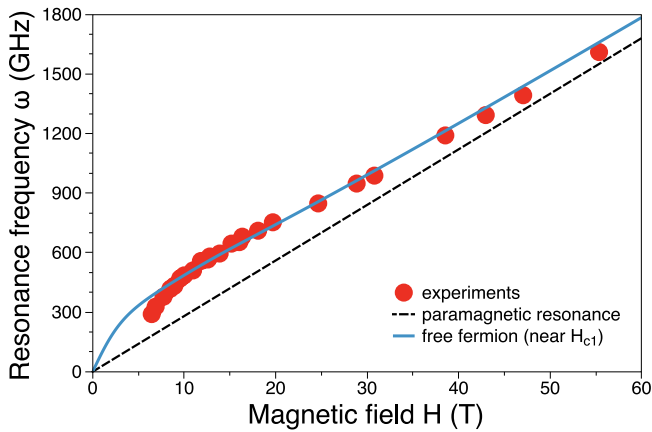}
 \caption{(Color online) Comparison of the free fermion theory
 \eqref{eq:YD_Hc1} with the experimental data on
 NDMAP at a temperature $T = 0.05J$.~\cite{kashiwagi_ndmap}
 The solid curve is the free fermion result \eqref{eq:YD_Hc1} near
 $H_{c1}$ and the solid circles denote the experimental data.
 We used the parameters $J = 30.0$ K, $D/J = 0.25$, and $g_e = 2.10$.
 The magnetic field is applied along $c$ axis, which corresponds
 to $z_p = 1$, $z_q=z_r = 0$.
 The dashed curve is high-temperature paramagnetic resonance frequency
 $\omega = g_e \mu_BH$.
 } 
 \label{fig:ndmap}
\end{figure}

We apply our theory of ESR shifts to an $S=1$ HAF chain compound
$\mathrm{Ni(C_5H_{14}N_2)_2N_3(PF_6)}$ (abbreviated to
NDMAP~\cite{honda_ndmap,zheludev_INS_ndmap,zheludev_ndmap,kashiwagi_ndmap}).
There are several $S=1$ HAF chain compounds, for instance,
$\mathrm{Ni(C_2H_8N_2)_2(NO_2)ClO_4}$ (abbreviated to
NENP~\cite{renard_nenp}),
$\mathrm{Ni(C_9H_{24}N_4)(NO_2)ClO_4}$ 
(abbreviated to NTENP~\cite{Escuer_NTENP}) and
$\mathrm{Ni(C_5H_{14}N_2)_2N_3(ClO_4)}$ (abbreviated to
NDMAZ~\cite{honda_NDMAZ}).
Among these $S=1$ HAF chain compounds, NDMAP is most suitable to our
purpose because NENP has an effective staggered magnetic field 
$h\sum_j (-1)^j S^x_j$ and NTENP has a bond alternation $\delta \sum_j
(-1)^j \bm S_j \cdot \bm S_{j+1}$.
The staggered magnetization mixes the singlet ground state $|g\rangle$ and
the triplet excited states $|e\rangle$: $\langle e|\sum_j (-1)^j S^x_j
|g \rangle \not=0$.~\cite{sakai_nenp}
This mixing changes the selection rule of ESR.
Such an interaction is uncovered by our theory.
Although the bond alternation does not induce the mixing, 
when $H=0$, NTENP has a different ground state from that of
\eqref{eq:H_0th}.~\cite{suzuki_NTENP}
Recently, NTENP has been field theoretically analyzed by using a
sine-Gordon model.~\cite{tamaki_NTENP}
The compound NDMAZ has very similar crystal structure to NDMAP.
In fact, our theory is applicable to NDMAZ.
But, NDMAZ has stronger exchange interaction $J\approx 70.6$ K
than NDMAP.
The large $J$ makes the experimental investigation of the field-induced
critical phase difficult because of the large $H_{c1}$.

Parameters of NDMAP are estimated as follows.~\cite{zheludev_INS_ndmap}
\begin{equation}
 J \approx 30.0 \, \mbox{K}, \quad
  D/J \approx 0.25
  \label{eq:parameters}
\end{equation}
The parameter $E$ is much smaller than $D$.
Here we consider the field orientation perpendicular to the easy plane,
$(z_p, z_q, z_r) = (1,0,0)$.
Thus, the normalized shift is independent of the anisotropy $E$.
\begin{equation}
 \omega_r = g_e\mu_B H -2D Y_D(T,H)
  \label{eq:freq}
\end{equation}
The Land\'e $g$ factor is $g_e = 2.11$.~\cite{kashiwagi_ndmap}
We substitute the free fermion theory near the lower critical field
\eqref{eq:YD_Hc1} into \eqref{eq:freq} and compare it with
experimental data by Ref.~\onlinecite{kashiwagi_ndmap} (Fig.~\ref{fig:ndmap}).
They show semiquantitative agreement.
Our theory gives a concrete support to the estimation
\eqref{eq:parameters}.

\begin{figure}
 \centering
 \includegraphics[width=\linewidth]{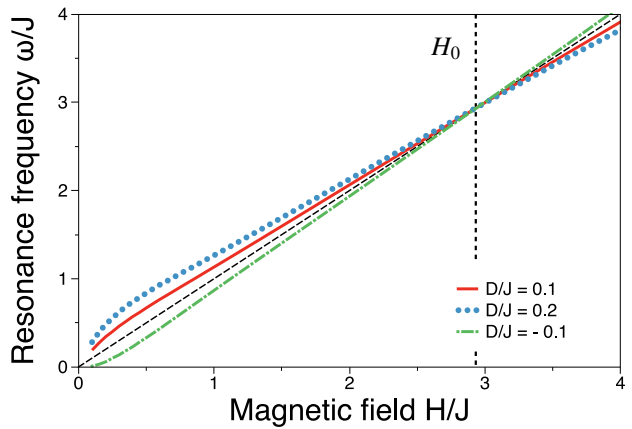}
 \caption{(Color online) QMC data of the resonance frequency
 \eqref{eq:freq} at $T=0.1J$.
 The $g$ factor is set to $g_e\mu_B = 1$ for simplicity.
 The dashed line $\omega = H$ corresponds to the paramagnetic resonance
 frequency. 
 Several cases $D/J = 0.1$, $0.2$ and $-0.1$ are shown.
 Note that there is a zero point $H_0 \sim 3J$
 where the shift \eqref{eq:NT} vanishes.
 }
 \label{fig:H0}
\end{figure}

Note that there is a zero point $H=H_0(T)$ where the ESR shift vanishes,
\begin{equation}
 Y_D(T,H_0(T))=0.
  \label{eq:def_H0}
\end{equation}
In addition to the trivial solution $H_0=0$,
at $T = 0.1J$, one can find a zero point $H_0 \sim 3J$  in Fig.~\ref{fig:H0}.
We show several cases $D/J = 0.1$, $0.2$ and $-0.1$ with the fixed $J$.
One will be able to experimentally observe the zero point $H=H_0$
if an $S=1$ HAF chain compound with smaller $J \lesssim 15$~K is found.
In general, the zero point $H_0(T)$ depends on the temperature $T$.
The non-trivial solution $H_0(T)$ of \eqref{eq:def_H0} exists in a wide range
of the temperature because $Y_D(T,H)$ is negative in $H \ll H_{c1}$
and positive in $H \sim H_{c2}$.
In contrast, as we will discuss in the Appendix, for the
exchange anisotropy
\begin{equation}
 \mathcal H' = \sum_j \sum_{a=p,q,r} J'_a S^a_j S^a_{j+1},
  \label{eq:XXZ}
\end{equation}
we find that the ESR shift in the first order of the
anisotropy does not change its sign in the entire range
of $H$.

By measuring the zero-field excitation gaps, the
symmetry of the Hamiltonian \eqref{eq:H_3} can be identified
experimentally.
Let us suppose that the Hamiltonian has a uniaxial, U(1) symmetry,
broken from the rotational SU(2) symmetry.
This is consistent with a presence of either the single-ion
anisotropy~\eqref{eq:SIA} or the exchange
anisotropy~\eqref{eq:XXZ}.
It is usually difficult to distinguish these two kinds of anisotropic
interactions because they often lead to
qualitatively the same consequences in observables.  However, the
presence or absence of the zero point $H_0(T)$ of the shift
at $T \sim H_{c1}$ is a clear signature which distinguishes
the two cases.
This may provide a new application of ESR, which possesses a high
sensitivity to anisotropy unavailable in other types of
measurements.

\section{\label{sec:summary}Summary}

We theoretically investigated the ESR shift caused by a weak
single-ion anisotropy in the $S=1$  HAF chain.
We applied the Kanamori-Tachiki theory \eqref{eq:NT} to this system,
and analyzed it both analytically and numerically.
The formula \eqref{eq:NT} is factorized to $\delta \omega = f_D(\bm z)Y_D(T,H)$,
which is composed of the $T$, $H$-independent geometrical
factor $f_D(\bm z)$ and the $T$, $H$-dependent factor $Y_D(T,H)$.
In this paper we call $Y_D(T,H)$ as the normalized ESR shift because
the factor $f_D(\bm z)$ can be regarded as a constant if we fix the
field orientation $\bm z$.
In contrast, the normalized shift $Y_D(T,H)$ does not depend on the
field orientation.
Thus, this factorization allows the general analysis of the ESR shift
without specifying the parameters $D$ and $E$.

Quantum Monte Carlo calculations revealed non-monotonic magnetic field
dependence of the normalized shift $Y_D(T,H)$.
The field dependence reflects the finite-temperature crossover
of the $S=1$ HAF chain, the low-field gapped phase ($H<H_{c1}$),
the field-induced critical phase ($H_{c1} < H<H_{c2}$), and
the fully polarized phase ($H_{c2} < H$).
We employed several effective field theories to explain the field
dependence of $Y_D(T,H)$ in each phase.
We used the exact form factors to compute $Y_D(T,H)$ in the dilute limit
$H, T \to 0$.
We extend the result in the low-field limit to the finite-field region
$H \sim H_{c1}$ by replacing the FZ operators of the lowest excitations
to the fermionic creation and annihilation operators.
This replacement is reasonable in $H \lesssim H_{c1}$ and it
works quite well (Fig.~\ref{fig:YD_Hc1}).
Above $H_{c1}$, the system is regarded as the TL liquid.
Although we did not go into detail of the ESR shift of the TL liquid
in the field-induced critical phase, it can be extracted from the
analyses around $H_{c1}$ and $H_{c2}$.
Near the upper critical field $H_{c2}$, the free fermion analysis
is again effective (Fig.~\ref{fig:YD_Hc2}).

Our analysis is found to agree semiquantitatively with the experimental
data of NDMAP in Ref.~\onlinecite{kashiwagi_ndmap}.  Our theory
correctly reproduces the approaching of the resonance frequency to the
paramagnetic resonance frequency $\omega = g_e\mu_BH$.  Furthermore, we
predicted the existence of the special value $H_0$ of the magnetic field
where the ESR shift vanishes $\delta\omega = 0$.
Such a sign change is absent in the case of an exchange anisotropy.

As a final remark, we point out that one can experimentally determine 
the field dependence of nontrivial quantities such as $\langle (S^z_j)^2
\rangle$ and $\langle S^z_j S^z_{j+1}\rangle$, from the ESR shifts~\eqref{eq:YD} and \eqref{eq:YJ'}.
The quantity $\langle (S^z_j)^2 \rangle$ is a nontrivial function of $H$ and $T$:
in an isotropic chain, it takes $2/3$ at $H=0$, decreases first as $H$
is increased, but increases asymptotically towards the saturation
value $1$ in the limit $H \to +\infty$.
This non-monotonic dependence is reflected in the shift \eqref{eq:YD}.

\section*{Acknowledgments}

This work is supported by Grant-in-Aid for Scientific Research
No.~21540381 (M.O.) and the Global COE Program ``The Physical
Sciences Frontier'' (S.C.F.) from MEXT, Japan.
We thank the ALPS project for providing the quantum Monte Carlo code.~\cite{ALPS1}

\appendix

\section{\label{app:exchange}Exchange anisotropy}

\begin{figure}[t!]
 \centering
 \includegraphics[width=\linewidth]{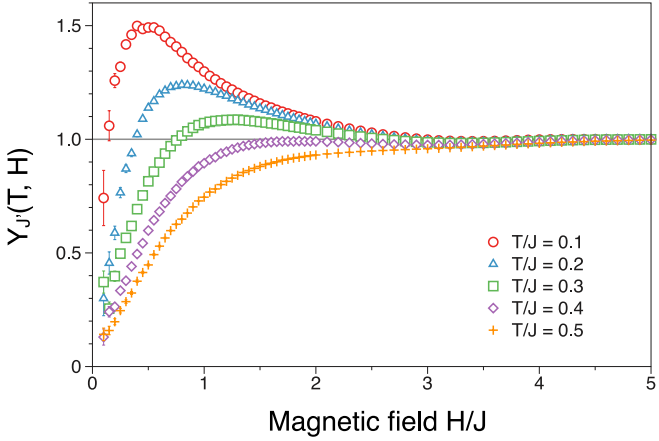}
 \caption{(Color online)
 Quantum Monte Carlo results of the normalized ESR shift
 \eqref{eq:YJ'} induced by the exchange anisotropy \eqref{eq:XXZ}
 for temperatures $T/J = 0.1$ -- $0.5$.
 The system size is $L=40$ sites.
 The maximum around $H = H_{c1} + T$ is also found in this case.
 The field dependence of $Y_{J'}(T,H)$ in a lower field region $H
 < J$ at low temperatures $T< 0.3J$ looks similar to that of
 $Y_D(T,H)$ except the overall sign, $Y_{J'}(T,H) \propto -Y_D(T,H)$.
 In a relatively higher temperature $T>0.4J$, the non-monotonic behavior
 of the normalized shift vanishes.
 }
 \label{fig:YJ'}
\end{figure}

\begin{figure}
 \centering
 \includegraphics[width=\linewidth]{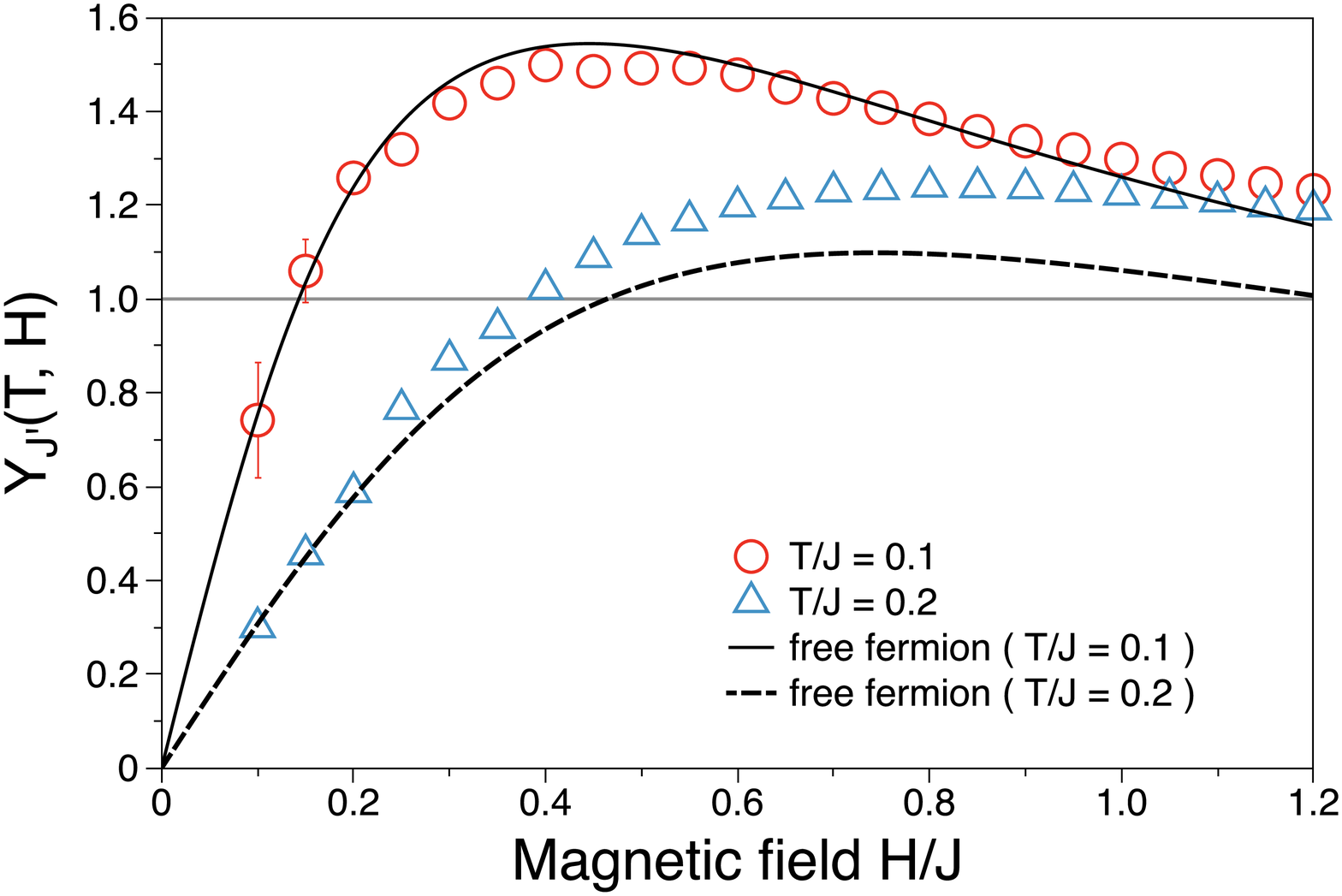}
 \caption{(Color online)
 Comparisons of QMC and \eqref{eq:YJ'_Hc1} at $T/J = 0.1$ and $0.2$.
 The open circles ($T/J = 0.1$) and triangles ($T/J = 0.2$) denote the
 QMC results.
 The solid (dashed) curve represents \eqref{eq:YJ'_Hc1} with $Z'_2=0.41$ at $T/J
 = 0.1$ ($T/J = 0.2$).}
 \label{fig:YJ'_Hc1}
\end{figure}

We have so far considered the single-ion anisotropy as a resource of
anisotropic interactions.
In this appendix, we treat a perturbative exchange anisotropy
instead of the single-ion anisotropy \eqref{eq:SIA}.
The ESR shift caused by the exchange anisotropy \eqref{eq:XXZ}
is also factorized just like \eqref{eq:dw_D}:
\begin{align}
 \delta \omega &= f_{J'}(\bm z) Y_{J'}(T,H)
 \label{eq:dw_J'} \\
 f_{J'}(\bm z) &= \sum_{a=p,q,r} J'_a (1-3{z_a}^2)
 \label{eq:fJ'} \\
 Y_{J'}(T,H)
 &= \frac 1{2\langle S^z \rangle_0} \sum_j \sum_{a=x,y,z}
 (3\delta_{az}-1)\langle S^a_j S^a_{j+1} \rangle_0
 \label{eq:YJ'}
\end{align}
We compute the normalized shift \eqref{eq:YJ'} by QMC
in the same manner as \eqref{eq:YD}.
Fig.~\ref{fig:YJ'} shows QMC results of the normalized shift
\eqref{eq:YJ'} at temperatures $T/J = 0.1$ -- $0.5$.
The normalized shift $Y_{J'}(T,H)$ behaves similarly to $Y_D(T,H)$
in a region where $T< 0.3J$ and $H<J$ hold.
On the other hand, in a higher field region $H>J$, the normalized
shift quickly saturates to 1.

First we consider the zero-field case.
The effective field theory O(3) NLSM works well at $H=0$.
When we move on to the continuum limit,
we approximate the product $S^a_j S^a_{j+1}$ by the composite
operator $[S^a(x)]^2$:
\begin{equation}
 S^a_j S^a_{j+1} \sim -C [S^a(x)]^2.
  \label{eq:SS}
\end{equation}
The coefficient $C$ is a non-universal constant.
When keeping only the most relevant term $S^a(x)S^a(x+a_0) \sim - S(S+1)
n^a(x)n^a(x+a_0)$, we may assume $C>0$ because the field $\bm n(x)$ is
smoothly varying on $x$.
Here $a_0$ is the lattice spacing and set to unity.
The replacement \eqref{eq:SS} immediately leads to
$Y_{J'}(T,H) \propto -Y_D(T,H)$ in the infinitesimal field region $H \ll H_{c1}$.
This relation is consistent with numerical
results (Figs.~\ref{fig:qmc} and \ref{fig:YJ'}).

Next, we extend our discussion to the finite field region $H \sim H_{c1}$
in the exactly same manner with Sec.~\ref{sec:Hc1}.
We assume that the replacement \eqref{eq:SS} is also valid under
not so weak magnetic field $H \sim H_{c1}$.
Then, the normalized shift $Y_{J'}(T,H)$ near $H=H_{c1}$ is given by
\begin{equation}
 Y_{J'} (T,H) = -\frac{3Z'_2}{2m(T,H)} \int_{-\infty}^\infty
  \frac{dk}{2\pi} \bigl[2f_0(k) - f_+(k) - f_-(k)\bigr].
  \label{eq:YJ'_Hc1}
\end{equation}
We determine the phenomenological parameter $Z'_2$ by fitting
\eqref{eq:YJ'_Hc1} with QMC data at $T/J = 0.1$.
The fitting leads to $Z'_2 \approx 0.41$.
Fig.~\ref{fig:YJ'_Hc1} shows the comparison of \eqref{eq:YJ'_Hc1} and
QMC data for $T/J = 0.1$ and $0.2$.
The formula \eqref{eq:YJ'_Hc1} reproduces the QMC data well.
But, their agreement rapidly becomes worse as the temperature rises.
This discrepancy stems from the saturation value
$Y_{J'}(T,H)\to 1$ in the limit $H \to +\infty$.
While $Y_D(T,H)$ is negative in the low-field region $H \ll H_{c1}$,
$Y_{J'}(T,H)$ is positive there.
Thus, the sign change of the normalized shift does not
occur for $Y_{J'}(T,H)$.
As we have discussed in Sec.~\ref{sec:ndmap},
this is in contrast to the behavior of $Y_D(T,H)$,
which universally shows a sign change.

\end{document}